\documentclass[fleqn,12pt,twoside]{article}
\usepackage{espcrc1}

\usepackage{graphicx}
\usepackage[figuresright]{rotating}

\newcommand{\AmS}{{\protect\the\textfont2
  A\kern-.1667em\lower.5ex\hbox{M}\kern-.125emS}}
\newcommand\beq{\begin{equation}}
\newcommand\eeq{\end{equation}}
\newcommand\beqa{\begin{eqnarray}}
\newcommand\eeqa{\end{eqnarray}}
\newcommand\br{{\bf r}}
\title{Hadron-quark mixed phase in neutron stars}

\author{T. Tatsumi\address[MCSD]{Department of Physics, Kyoto University,
        Kyoto 606-8502, Japan}%
        \thanks{E-mail address: tatsumi@ruby.scphys.kyoto-u.ac.jp},
        M. Yasuhira\addressmark[MCSD]$^,$\address[YITP]{Yukawa Institute for
        Theoretical Physics, Kyoto University, Kyoto 606-8502, Japan}
        and
        D.N. Voskresensky\addressmark[YITP]$^,$\address[GSI]{Gesellschaft
        f\"ur Schwerionenforschung (GSI), Planck
        Str 1, D-64291 Darmstadt, Germany}$^,$\address[MIPE]{Moscow Institute
        for Physics and Engineering,
        Kashirskoe sh. 31, Moscow 115409, Russia}}
\begin{document}

\maketitle

\begin{abstract}
Possibility of structured mixed phases at first order  phase
transitions  is examined by taking into account of charge
screening and surface effects. Hadron-quark phase transition in
dense neutron star interiors is considered
 as a concrete example.
\end{abstract}

\section{INTRODUCTION}

Recently many efforts have been made for the derivation of the
equation of state (EOS) of the neutron star matter exhibiting
possibilities of various phase transitions in neutron star
interiors. Some of these possible phase transitions are the first
order phase transitions (FOPT) essentially affecting the EOS.
We will discuss this issue by considering the hadron-quark (H-Q)
matter phase transition\cite{vyt}.

In Ref. \cite{gle} Glendenning has pointed out that the usual
Maxwell construction is applicable only for systems with one
particle species and respectively one chemical potential, whereas
in neutron stars there are two relevant quantities, the charge and
baryon number chemical potentials. He has advocated to use the
Gibbs conditions and  demonstrated the appearance of a wide region
of the structured mixed phase, consisting of droplets, rods or
slabs of one phase embedded into the other phase, instead of the
Maxwell construction. Finite size effects, the
Coulomb interaction and the surface tension,
 in the treatment of the
mixed phase were disregarded. His idea was further exploited by
many authors. It was concluded that a wide region of the mixed
phase is inevitable and there should be no region of constant
pressure, opposite to that follows from the Maxwell construction.

On the other hand, Heiselberg et al. pointed out the importance of
the inclusion into consideration of the finite size effects 
to realize the
structured mixed phase \cite{hei}. They demonstrated that for
rather large values of the surface tension the mixed phase
constructed of rather small size droplets  becomes energetically
unfavorable compared to the two bulk-phase structure given by the
Maxwell construction. Recently it was argued that the Gibbs
condition of the equality of the electron chemical potentials is
not satisfied at the interface between the color-flavor-locked
phase (no electrons) and the nucleon matter phase and it is
described by the Maxwell construction \cite{alf}. In the above
treatments the charged particle density profiles are uniform in
each phase and furthermore corrections due to charge screening
effects are disregarded. In spite of the question on the role of
the screening  has been raised long ago, up to now there was no
consistent treatment of these inhomogeneity effects. Therefore, a
further study of the screening and surface effects seems to be of
prime importance.

\section{DENSITY FUNCTIONAL METHOD}

Consider a quark droplet (I) immersed in the ($n,p,e$) nuclear matter (II)
within the Wigner-Seitz cell. The narrow boundary region between them is
approximated by the sharp boundary $\partial D$. Then the total 
energy functional $E$ is
given by a functional of particle densities $\rho_i$\cite{par}:
\beq 
E[\rho(\br)]=\int_{D^{\rm I}} d^3r \epsilon^{
I}_{kin+str}(\rho_i^I(\br))+\int_{D^{\rm II}} d^3r \epsilon^{
II}_{kin+str}(\rho_i^{II}(\br)) +\int_{\partial
D}d^2r\epsilon_{S}+E_V, 
\eeq 
where the first and second terms are
the kinetic plus strong interaction contributions in both phases,
the third term is the surface energy contribution. The last term
is the usual Coulomb interaction energy, \beq
E_{V}=\frac{1}{2}\int\int d^3r d^3r'\sum_{i,j}
\frac{Q_i\rho_i(\br)Q_j\rho_j(\br')} {|{\bf r}-{\bf r'}|}, \eeq
with $Q_i$ being the particle electric charge. Then the generating
functional under the constraint of the particle number
conservation is introduced via the Legendre transformation, 
\beq
\Omega[\rho(\br)]=E[\rho(\br)]-\sum_i\mu_i^I\int_{D^{\rm I}}d^3r\rho_i^I(\br)
-\sum_i\mu_i^{II}\int_{D^{\rm II}}d^3r\rho_i^{II}(\br). 
\eeq Note that
$\Omega[\rho(\br)]$ is the analog of the thermodynamic potential
and the parameters $\mu_i^\alpha$ are the chemical potentials.
When we consider two conservation laws relevant in the mixed phase
: baryon number and charge conservation, these quantities
are well defined over the Wigner-Seitz cell, not restricted to each
domain. Accordingly the baryon number and charge chemical
potentials ($\mu_B$ and $\mu_Q$) , being linear combinations of
$\mu_i^\alpha$,
become constants over the whole space,
\begin{equation}\label{gibbs1}
\mu^{I}_B=\mu^{II}_B\equiv \mu_B,~~~\mu^{I}_Q=\mu^{
II}_Q \equiv \mu_Q;
\end{equation}
e.g. $\mu_Q=\mu_e^\alpha$ and $\mu_B=\mu_n =2\mu_d +\mu_u$.

Equations of motion are given by the variational principle: \beq
\frac{\delta\Omega[\rho(\br)]}{\delta\rho_i^\alpha(\br)}=0~~~{\rm
or}~~~ \mu_i^\alpha=\frac{\partial\epsilon_{kin+str}^\alpha}
{\partial\rho_i^\alpha}- (Q_i^{ \alpha}/e)V^\alpha
(\br)
, \,\,\, \alpha =\{I,II\},
\label{motion} \eeq
where $V^\alpha(\br)$ is 
the electric potential, 
\beq V=-\int
d^3r'\sum_{l}\frac{eQ_l\rho_l}{|{\bf r}-{\bf r'}|} =\left\{
\begin{array}{ll}
V^I(\br) & {\br}\in D^I\\
V^{II}(\br) &{\br}\in D^{II},
\end{array}
\right. \eeq 
and satisfies  the Poisson equation, \beq \Delta
V^\alpha=4\pi e\sum_{i}Q_i\rho_i^\alpha. \eeq Note here that the
electric potential can be shifted by an arbitrary constant, $V^0$,
and theory should be invariant under the redefinition of chemical
potentials in a proper way; 
$
V\longrightarrow V-V^0,
\mu_i^\alpha\longrightarrow \mu_i^\alpha+(Q_i^{ \alpha}/e)V^0.
$ 
One convenient choice is $V^0=0$ that means that the
electron density can be expressed as
$\rho_e=(\mu_e-V)^3/(3\pi^2)$. Another convenient choice is
$V^0=-\mu_e$ corresponding to $\rho_e=-V^3/(3\pi^2)$, see
\cite{vyt}.


If we expand the charge density in the r.h.s. of the Poisson
equation with respect to $V^\alpha$ around a reference
 value $V_{r}^\alpha$ up to the second order, we arrive at  the
equation \beq \Delta {\tilde
V}^\alpha=4\pi\sum_{j}Q_j^\alpha\rho_j^\alpha
(V^\alpha=V_r^\alpha)+\kappa^2_\alpha {\tilde V}^\alpha \eeq with
${\tilde V}^\alpha=V^\alpha-V_r^\alpha$. The coefficient
$\kappa_\alpha$ is the inverse of the Debye screening length, \beq
\kappa^2_\alpha\equiv 1/(\lambda^{\alpha}_D )^2=
\left.4\pi\sum_{i,j}Q_j^\alpha
Q_i^\alpha\frac{\partial\rho_i^\alpha}
{\partial\mu_j^\alpha}\right|_{V^\alpha=V_r^\alpha}. \eeq
Electrons (quarks) can be treated as having the uniform charge
density profile only if the droplet size and the inter-droplet
distance are essentially smaller than the screening length
$\lambda_{e,D}$ ($\lambda_{q,D}$). Otherwise, as we shall
demonstrate it below, the screening effects become essential and
the charged density profiles are not uniform. Estimates show that
the screening length for electrons $\lambda_{e,D}\geq 13~ \rm fm$
is longer  than that for the quarks, $\lambda_{q,D}\simeq 5~{\rm
fm}$.

Pressure in each phase is constant owing to the equation of motion
(\ref{motion}), 
and the pressure
balance condition at the boundary $\partial D$ reads 
\beq
P^I=P^{II}+P_S, 
\label{press} 
\eeq where $P_S$ denotes the surface
contribution. In the following we assume the spherical geometry
and presence of the sharp boundary parameterized by the constant
surface tension $\sigma$. Then Eq.~(\ref{press}) is equivalent to
the extremum condition,
$\partial\Omega/\partial R=0,
$
where $R$ is the droplet radius. Eqs.~(\ref{gibbs1}), (\ref{press}) are 
manifestation of the Gibbs conditions.

\section{RESULTS AND DISCUSSION}

In Fig.~1 we show the effective energy of the Wigner-Seitz
cell per droplet volume associated with the inhomogeneity of the
electric field profile,
$\delta\tilde\omega=3\Omega[\rho(\br)]/(4\pi R^3)$, up to the
second order in $\tilde V^\alpha$ as a function of the scaled
droplet radius  $\xi=R/\lambda^{\rm I}_D$.  The curves are presented
for two values of the concentration of phase I,  $f=(R/R_W)^3$,
$R_W$ is the radius of the Wigner-Seitz cell. Solid curves are for
$f=1/2$ and dashed curves are for a tiny value $f=0.01$. The
``e.m.'' curves show the partial contribution of the (screened)
electric field to the energy ($\propto \int (\nabla V)^2 d^3r$)
plus the surface energy. The difference between the non-labeled
curves and the ``e.m.'' curves shows the important role of the
secondary (correlation) effects coming from the dependence of the
bulk energies ($\epsilon_{kin+str}$) on the inhomogeneous
distribution of the electric field.  The ratio of the screening
lengths $\alpha_0 =\lambda_D^{ I}/\lambda_D^{II}$ is assumed to be
1.

We vary the value of the surface tension,
$\beta_1=3\sigma/(\lambda^{\rm I}\beta_0)$ with the energy scale
parameter $\beta_0$, admitting its uncertainties. E.g., for
typical values $\mu_{e}\simeq 170~$ MeV for the electron chemical
potential, $\mu_{n} \simeq 1020~$ MeV for the baryon (neutron)
chemical potential, $\alpha_c \simeq 0.4$ for the strong coupling
constant, and $m_s \simeq 120\div 150$~MeV for the strange quark
mass, we estimate $\beta_0 \simeq 1.6 m_{\pi}^4$. Then, with the
value $\sigma \simeq 1.3 m_{\pi}^3$ we obtain $\beta_1 \simeq 0.7
$, whereas with $\sigma \simeq 10~$ MeV/fm$^2 \simeq
0.14m_{\pi}^3$ we would get $\beta_1 \simeq 0.08$. The Coulomb
curves (labeled by ``C'') show the corresponding results obtained
neglecting screening effects.
We see that the ``C'' curves demonstrate pronounced minima at
$\xi=\xi_C\propto \beta_1^{1/3}$, while there is no minimum for
$\beta_1>\beta_{1c}$ if one includes screening effects. According
to our figure $\beta_{1c}\simeq 0.6$ (would be $\beta_{1c}\simeq
0.03$ for "e.m." curves, when correlation energy effects are
disregarded). Only for $\beta_1 \leq 0.01$ the minima at the "C"
curves differ not essentially from the minima at other curves.
Hence our results clearly show that the screening effects are
important for all realistic values of $\sigma$ and that the
structured mixed phase is proved to be prohibited due to the
kinetic instability of the droplets induced by the screening
effects provided  the surface tension is  larger than a critical
value. In absence of the mixed phase our charged distributions
describe the boundary layer between two separated phases existing
within the double-tangent (Maxwell) construction. Consideration of
non-spherical droplets (rods and slabs) does not change our
conclusions\cite{vyt}.
\begin{figure}[h]
\begin{center}
\includegraphics[scale=0.5]{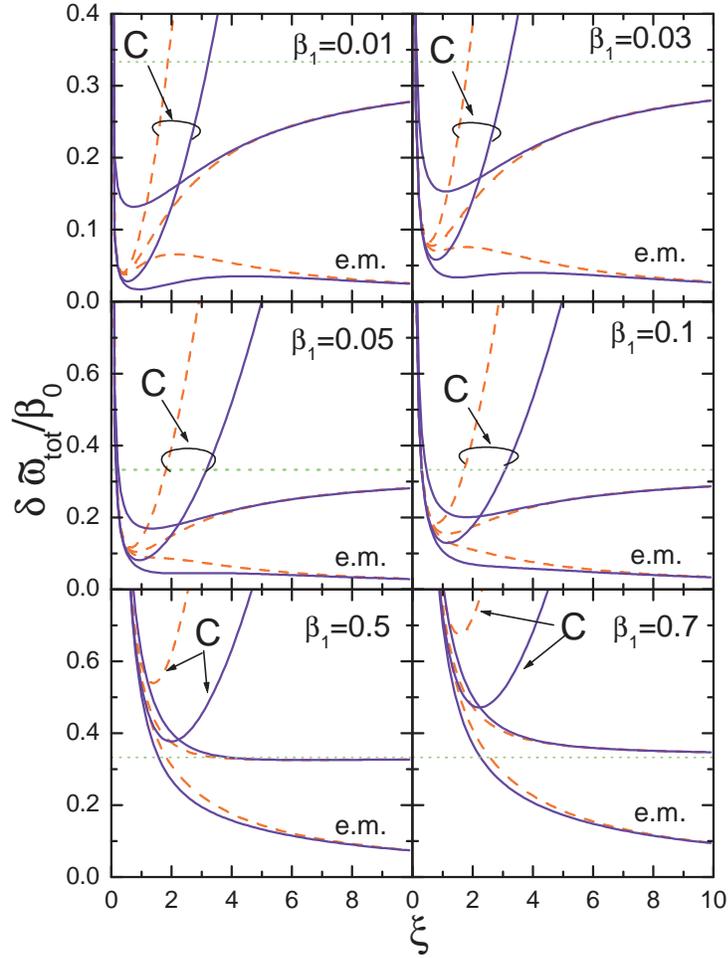}
\caption{Contribution to the effective energy per droplet volume
due to inhomogeneous charge distributions  versus scaled droplet
radius for the droplet concentration $f=1/2$ (solid curves) and a
tiny concentration $f=0.01$ (dashed curves), $\beta_0$ is the
scale-parameter for the energy, and $\beta_1$ is the parameter of
surface tension (see the text). The "C" curves are calculated
ignoring screening effects, the "e.m." curves include the electric
field energy (with screening effect included) plus surface energy,
ignoring however the correlation effects. }\label{fig2}
\end{center}
\end{figure}

\end{document}